# Possible Detection of Gamma Ray Air Showers in Coincidence with BATSE Gamma Ray Bursts

T. F. Lin, J. Carpenter, S. Desch, J. Gress, J. Poirier, and A. Roesch
*Physics Department, 225 NSH, University of Notre Dame, Notre Dame, IN 46556, USA*

**Abstract**

Project GRAND presents the results of a search for coincident high-energy gamma ray events in the direction and at the time of nine Gamma Ray Bursts (GRBs) detected by BATSE. A gamma ray has a non-negligible hadron production cross section; for each gamma ray of energy of 100 GeV, there are 0.015 muons which reach detection level (Fasso & Poirier, 1999). These muons are identified and their angles are measured in stations of eight planes of proportional wire chambers (PWCs). A 50 mm steel plate above the bottom pair of planes is used to distinguish muons from electrons. The mean angular resolution is $0.26^o$ over a $\pm 61^o$ range in the XZ and YZ planes. The BATSE GRB catalogue is examined for bursts which are near zenith for Project GRAND. The geometrical acceptance is calculated for each of these events. The product is then taken of the GRB flux and GRAND's geometrical acceptance. The nine sources with the best combination of detection efficiency and BATSE's intensity are selected to be examined in the data. The most significant detection of these nine sources is at a statistical significance of $+3.7\sigma$; this is also the GRB with the highest product of GRB flux and geometrical acceptance.

## 1  Introduction:

Gamma Ray Bursts (GRBs) are an interesting physics phenomenon which have been around for quite some time; as yet there is no consensus explanation for GRBs. Whatever the GRB source environment, however, it seems quite likely that the burst environment involves collisions with ultra relativistic ejecta (Paczynski, 1986; Goodman, 1986). Any baryons present in this ejecta could be accelerated to very high energies (Waxman, 1995). Secondary interactions of such baryons would then produce energetic gamma rays in coincidence with gamma ray bursts. Detection of such high energy photons would provide important insights into the GRB source environment, which is the motivation for the present search. It has been suggested (Mannhelm et al., 1996) that the GRB rate for threshold energy larger than 200 GeV is 10 GRBs per year. Project GRAND will have an even higher GRB observed rate because of GRAND's lower threshold energy (10-30 GeV). Project GRAND utilizes the fact that gamma rays have a detectable signal from their hadron production cross section which produces pions which can then decay to muons. A Monte Carlo calculation for this probability is presented in a paper to this conference (Fasso & Poirier, 1999). Such muons then have a good chance to reach Project GRAND where they are identified as muons and their angles are measured.

## 2  Experimental Array:

Project GRAND is located just north of the University of Notre Dame campus approximately 150 miles east of Chicago and 220 m above sea level at $86^o$ W and $42^o$ N. It detects cosmic ray secondaries at ground level by means of 64 tracking stations (huts) of proportional wire chambers (PWCs). Each hut is 2.4 m x 2.4 m x 1.5 m. In each station there are four chambers each containing two x-y planes (Linsley et al., 1987; Poirier et al., 1990) arranged vertically spaced above each other. Each secondary particle track is measured to $0.26^o$ absolute precision (average value in each of two orthogonal planes) (Gress et al., 1991). Each plane has an active area of 1.25 $m^2$ with 80 detection cells, 10 mm high and 14 mm wide which gives

Project GRAND a total detection area of 80 m². A 50 mm steel plate inserted between the third and fourth PWCs allows each track to be identified as a muon (or not); 4% of electron tracks are misidentified as muon tracks and 96% of muons are identified as muons tracks. Since for single track data there are about four times as many muon tracks as there are electron tracks, then 99% of the single tracks which pass the muon algorithm are muons (Gress et al., 1990).

These secondary muons are the result of interactions of the primary gamma rays with the atmosphere producing pions, which then decay to muons. The pions are produced at a small angle relative to the primary gamma rays and decay into muons at a small angle relative to the parent pion. These muons are then bent by the earth's magnetic field and scattered in the earth's atmosphere resulting in an effective resolution (including all these effects) for the primary cosmic ray of ± 5° (in each of the two orthogonal planes). The muon threshold energy is 0.1 GeV for vertical tracks, increasing at approximately $1/\cos(\theta)$ for $\theta$-angles inclined from vertical. The current rate for recording and identifying muons is 2000 Hz. In the past two years, 100 billion such muons have been collected with their angles and times measured and reduced to solar and stellar angle coordinates. Data previous to this time were taken with a smaller detector array during Project GRAND's construction and had a correspondingly smaller counting rate and more intermittent running.

## 3   Data Analysis:

The complete GRB table, the flux table, and the duration table are downloaded from the BATSE website (http://www.batse.msfc.nasa.gov/batse/). These tables provide the following information for each GRB: Universal Time (UT), right acension (RA,$\alpha$), declination (DEC,$\delta$), Flux, ST90 (start time for the T90 interval relative to the UT time), T90 (the duration of time after ST90 which includes 90% of the total observed counts in the GRB), etc. RA, DEC, and UT information are then transformed to local Altitude (Alt) and Azimuth. Using the altitude information, Project GRAND detector's acceptance (see Figure 1) is calculated for each GRB:

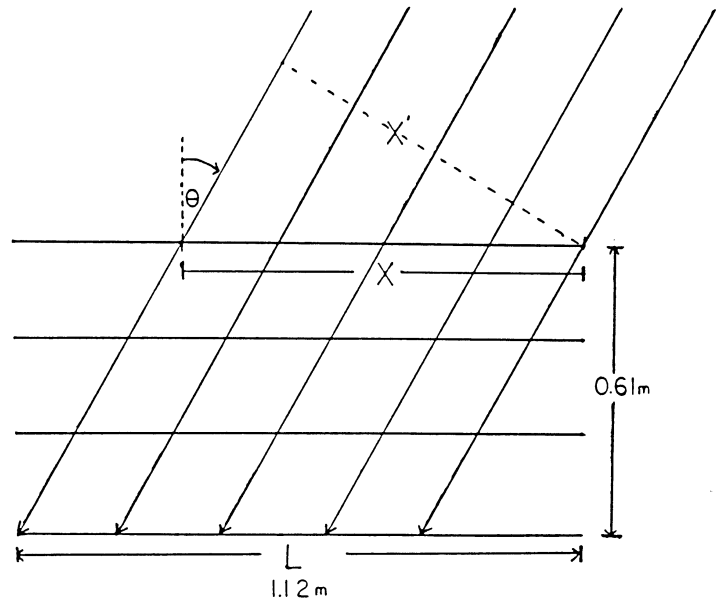

**Figure 1:** Plan of experiment to calculate Project GRAND's geometric acceptance as a function of angle from zenith, $\theta$.

$$\text{Acceptance} = (X / L) * \cos(\theta) * \cos^2(\theta)$$

The ratio of X and L from Figure 1 represents the reduced effective area for the detector for a muon track at an angle θ from zenith. This ratio is multiplied by cos(θ) to project the length X normal to the muon track, X'. The angular distribution of muons from zenith angle is proportional to $\cos^2(\theta)$ (Particle Data Group, 1998) which takes into account the increased thickness of the atmosphere for tracks inclined from the zenith. The likelihood of Project GRAND to observe each GRB is proportional to the product of GRAND's acceptance and the GRB's flux as measured by BATSE. After sorting these probabilities, the 20 most likely GRB candidates were selected. For those 20 GRBs, data for 10 of them were located on archived data tapes. Of those 10 GRBs, one was found to have three stations with large time dependent inefficiencies thus creating artificial muon count deviations during the time interval of the GRB and so it was discarded.

A window with ± 5° in DEC and ± 5°/cos(δ) in RA was centered on the location of the GRB as provided by BATSE. The division by the cosine of the DEC in RA is because the width of RA to retain an equivalent area increases with increasing DEC. The muon accumulation in that window was studied with a histogram of total counts versus time (1 bin per sec) which covered 1000 sec before the burst and 1000 sec after the beginning of the GRB with each histogram covering 100 sec. The total entries in each histogram increase (or decrease) with time due to the detector's rotation toward (or away) from the location of the GRB. All histograms except the histogram containing the GRB (histo-GRB) are treated as background. A graph of total muon counts vs. histogram number (histo-time) was plotted and a linear fit was made to estimate the background based on assuming the background to be linear with time variations. The start time of each GRB was the sum of the UT time and the ST90 time, both provided by BATSE. An error of one second was assigned to the time synchronization between BATSE and GRAND. Therefore, one second was added to the beginning and end of each GRB; the time was then truncated to the nearest second. The average background was taken as the value at the mid time of the GRB according to the background's linear fit. The signal calculated for the GRB is the difference between the total counts during the T90 interval provided by BATSE and the background found from the linear fit. The error on the signal was taken to be the statistical error of this difference.

The results of these nine GRBs are listed in Table 1. Listed are the UT time of the burst (date and hour in universal time of their trigger), the actual burst start time in seconds from the UT time (ST90), the time in seconds for 90% of the burst's counts to occur (T90), the right ascension angle (RA), and the declination angle (DEC). Then the following are calculated: altitude angle (Alt), log (base 10) [GRAND's geometrical acceptance for the burst multiplied by BATSE's observed flux in their highest energy interval]=(Pred), background (Bckgrd), signal above background (Sig), error on signal (δSig), and signal divided by error on signal (Sig/δSig) to give the statistical significance of each burst's detection.

Table 1: The Results of Project GRAND

| UT (trigger) day/hr | ST90 | T90 | RA | DEC | Alt | Pred | Bckgrd | Sig | δSig | Sig/δSig |
|---|---|---|---|---|---|---|---|---|---|---|
| 11/10/97  18.8908 | 21.6 | 195.2 | 242 | 50 | 81 | 5.18 | 14021 | 445 | 120 | 3.7 |
| 05/26/94  20.3349 | 3.8 | 48.6 | 132 | 34 | 66 | 4.68 | 507 | 19 | 23 | 0.8 |
| 06/23/93  3.1595 | -1.5 | 1.4 | 262 | 46 | 69 | 4.05 | 37 | 11 | 7 | 1.6 |
| 04/20/98  10.1146 | 0.9 | 39.9 | 293 | 27 | 68 | 4.02 | 1751 | 86 | 43 | 2.0 |
| 04/28/96  13.2089 | 2.3 | 172.2 | 304 | 35 | 70 | 3.83 | 1937 | 45 | 45 | 1.0 |
| 01/05/98  0.7450 | 0.6 | 36.8 | 37 | 52 | 79 | 3.46 | 2136 | -103 | 45 | -2.3 |
| 06/08/93  19.6928 | -0.9 | 1.9 | 105 | 41 | 89 | 3.18 | 74 | -1 | 8.5 | -0.1 |
| 03/01/98  6.1163 | 0.3 | 36.0 | 148 | 35 | 76 | 3.17 | 2175 | 86 | 47.5 | 1.8 |
| 04/24/93  19.3167 | 0.1 | 19.3 | 54 | 42 | 88 | 3.17 | 338 | -25 | 17.7 | -1.4 |

The three GRBs with the highest Sig/δSig were tested to see if perhaps some of the huts produced artificial count abundances via noise or erratic time dependences. It was determined through a hut by hut count vs. time test that these huts exhibited no time structure other than statistical variations; only their sum provided the signals which are reported in Table 1.

## 4    Conclusion:

In nine attempts, six of the bursts had an abundance of counts and three of the bursts had a deficiency of counts. Five of the bursts were ≥ +1σ, two bursts were ≥ +2σ, and one burst was ≥ +3σ. From statistics, the expected counts were 1.4 events for ≥ +1σ, 0.21 events for ≥ +2σ, and 0.0018 events for ≥ +3σ. Statistically, the probability of detecting, in nine attemps: ≥ 5 events at ≥ +1σ is 0.8%; detecting ≥ 2 events for ≥ +2σ is 1.6%; and detecting ≥ 1 event at ≥ +3.7σ is 0.02%. These nine GRBs had a summed sigma of +7.1. The burst at +3.7σ is also the burst out of the nine that had the highest flux and was the burst predicted most likely to have been detected. In this sample of nine candidates, these features of the data suggest that Project GRAND has detected one or more gamma ray bursts in coincidence with BATSE's published times and angles.

Thanks to Grant Mathews for theoretical discussions and suggestions for this paper.

This research is presently being funded through grants from the University of Notre Dame and private donations.